\documentclass[conference,a4paper]{APSIPA2017}
\usepackage{multirow}
\usepackage{graphicx}
\usepackage{amsmath}
\usepackage[psamsfonts]{amssymb}
\usepackage{amsxtra}
\usepackage{threeparttable}

\usepackage{amsmath,amssymb}

\usepackage{txfonts}
\usepackage{bm}

\begin{document}

\title{Cross-lingual Speaker Verification with Deep Feature Learning}

\author{%
\authorblockN{%
Lantian Li\authorrefmark{2}, Dong Wang\authorrefmark{2}, Askar Rozi\authorrefmark{2}, Thomas Fang Zheng\authorrefmark{2}
}

\authorblockA{%
\authorrefmark{2}
Center for Speech and Language Technologies, Research Institute of Information Technology \\
Department of Computer Science and Technology, Tsinghua University, China
}


%
}

\maketitle
\thispagestyle{empty}

\begin{abstract}

Existing speaker verification (SV) systems often suffer from performance degradation
if there is any language mismatch between model training, speaker enrollment, and test. A major
cause of this degradation is that most existing SV methods rely on a probabilistic model
to infer the speaker factor, so any significant change on the distribution of the speech signal
will impact the inference. Recently, we proposed a deep learning model that can
learn how to extract the speaker factor by a deep neural network (DNN). By this feature learning,
an SV system can be constructed with a very simple back-end model. In this paper, we investigate
the robustness of the feature-based SV system in situations with language mismatch. Our experiments
were conducted on a complex cross-lingual scenario, where the model training was in English, and
the enrollment and test were in Chinese or Uyghur. The experiments demonstrated that the feature-based
system outperformed the i-vector system with a large margin, particularly with language mismatch
between enrollment and test.


\end{abstract}

\section{Introduction}
\label{sec:intro}

Speaker verification (SV) is an important biometric recognition technology and has gained great popularity in a wide range of applications,
such as access control, transaction authentication, forensics and personalization.
After decades of research, speaker verification has gained significant performance improvement, and has been deployed in some practical applications~\cite{campbell1997speaker,reynolds2002overview,kinnunen2010overview,hansen2015speaker}.
In spite of the great achievement, current speaker verification systems perform well only if the enrollment and test utterances are well matched, otherwise the performance will be seriously degraded.

Language discrepancy is a particularly mismatch that degrades SV performance.
In the latest decades, the development of the Internet has greatly sped up information exchange
and has glued the entire world together. An interesting consequence is the widespread bilingual and multilingual phenomena. This leads to the notorious cross-language issue for SV systems. There are two types of language mismatch: the first
type of mismatch occurs when the system is trained with data in one language, but operates in other languages; the second type of
mismatch occurs when a speaker enrolles in one language but tests in a different language. These two types of mismatch may be
mixed, i.e., the training, enrollment, and test are in three different languages.
As an example, in Urumchi, a large city in the western China, Uyghur and Chinese are both official
languages and are used by many people interchangeably even without notice. Hence for people living Urumchi, language mismatch between enrollment and
test is very likely to occur. However, for both Chinese and Uyghur, there are no standard databases to train the SV system, and we have to
resort to speech databases in English, e.g., the Fisher database published by LDC. This practical scenario involves two types of
language mismatch, and is highly complex and challenging for SV systems.

Intuitively, the mismatch in language should not be a problem, since a person's speaker trait is nothing to do with
what he/she is speaking. From this perspective, SV should be naturally cross-lingual. However, for most present SV systems,
the language mismatch is indeed a serious problem and great performance degradation has been reported~\cite{ma2004english,auckenthaler2001language,Abhinav,askar2016}.
We argue that a deep reason for the cross-lingual loss is the probabilistic modelling approach used by most of the existing SV systems,
particularly the famous Gaussian mixture model-universal background model (GMM-UBM)~\cite{Reynolds00} architecture,
and the succeeding joint factor analysis (JFA) model~\cite{Kenny07} and its `simplified' version, the state-of-the-art i-vector model~\cite{dehak2011front}. By these models, the speaker property is modeled as an `additive component'
augmented to the basic phonetic variation that is represented by the Gaussian components, e.g., a mean vector shift from the
speaker independent Gaussian. This implies that if the distribution of the basic phonetic variation is changed (the case with
cross-lingual operation, both enrollment and test), the speaker property will be poorly represented. This is why existing
SV systems perform bad with language mismatch.


Perhaps the best solution for the cross-lingual problem is to retrieve high-quality speaker features. If
the phonetic variation can be effectively eliminated from the feature, we can entirely discard the probabilistic
model such as the GMM, and the SV model will be ideally language independent.
Essentially, if we can find such a feature, most of the difficulties of the existing SV models will be solved, not limited to the
cross-lingual problem. This has motivated quite some researchers to pursue `fundamental' speaker features, e.g.,~\cite{Kinnunen10,wang2013vocal}.
However, most of the feature engineering methods rely on human knowledge, which turns out to be extremely difficult. This is why the state-of-the-art SV systems are still based on probabilistic models.

Fortunately, we recently found a powerful \emph{feature learning} approach based on a
deep neural network (DNN) structure~\cite{li2017deep}.
This structure consists of a convolutional (CN) component and a time-delay (TD) component,
designed to learn local patterns and extend the temporal context, respectively.
Our experimental results demonstrated that this CT-DNN model could learn strong speaker sensitive
features and outperformed the i-vector model especially in short utterance conditions.
By this deep feature learning structure, the speaker-discriminative information can be preserved and
strengthened, while speaker-irrelevant variations, especially the phone content, are diminished and removed.
We conjecture that this feature learning SV approach is particularly robust in scenarios where
there is significant phonetic variation, for example in the cross-lingual circumstances.

In this paper we will investigate the performance of the feature learning SV approach with a cross-lingual SV task,
where the model is trained with a large English speech database, and the enrollment and test could be in
Chinese or Uyghur. We found that the feature-based SV outperformed the state-the-art i-vector system with a
large margin. Besides, we designed a phone-aware deep feature learning structure
that further improves the cross-lingual SV performance.

The organization of this paper is as follows: we firstly describe some related work on the deep feature learning approach in Section~\ref{sec:rel},
and then present the CT-DNN structure in Section~\ref{sec:model}.
The experiments will be presented in Section~\ref{sec:exp}, followed by some conclusions and discussions in Section~\ref{sec:conl}.

\section{Related work}
\label{sec:rel}

Cross-lingual SV has been studied by some authors.
For example, Ma et al.~\cite{ma2004english} studied the enrollment-test mismatch and found that it
caused significant performance degradation for speaker recognition.
Auckenthaler~\cite{auckenthaler2001language} investigated the mismatch between training and operation,
within the GMM-UBM architecture.
They found considerable performance degradations if the speech data used to train the UBM
and the speech data used to enroll/test speakers were in different languages.
Abhinav et al.~\cite{Abhinav} studied the same problem within the state-of-the-art i-vector
architecture~\cite{dehak2011front},
and investigated both the training-operation mismatch and the enrollment-test mismatch. Their
results confirmed that language mismatch, despite where it occurs, leads to
significant performance degradation.

Some compensation methods have been proposed to alleviate the cross-lingual impact.
Akbacak et al.~\cite{akbacak2007language} proposed two normalization techniques:
normalization at the utterance-level via language identification and normalization at the segment-level via multilingual phone recognition.
Askar et al.~\cite{askar2015cross} applied the constrained maximum likelihood linear regression (CMLLR) to learn a
transform that maps acoustic features from one language to another.
Lu et al.~\cite{lu2009effect} formulated the cross-lingual problem in a more elegant Bayesian framework, and the joint
factor analysis (JFA) formulation was extended by adding a latent factor to represent the language. This language factor
was inferred and compensated for the enrollment and test. Recently, Askar et al.~\cite{askar2016} proposed a
phone-aware PLDA approach that involves phone information when training PLDA.

All the above compensation methods are based on the probabilistic modeling framework. There
are also some work on features. For example, Wang et al. ~\cite{wang2013vocal} studied vocal fold features,
such as the residual phase cepstral coefficients (RPCCs) and the glottal glow cepstral coefficients (GLFCCs).
They employed these features to address the cross-lingual challenge and found some reasonable performance improvement.

The idea of deep feature learning was originated by Ehsan et al.~\cite{ehsan14}.
They constructed a DNN model with $496$ speakers in the training set as the targets.
The frame-level features were read from the activations of the last hidden layer,
and the utterance-level representations (called `d-vector') were obtained by averaging over the frame-level features.
Although the pure d-vector system was not better than the state-of-the-art i-vector system, this work triggered
much interest on the deep neural approach for SV, though most followers focused on an end-to-end scheme ~\cite{heigold2016end,zhang2017end,snyderdeep16,li2017}
that learns a neural scoring network together with the feature network. This actually departed from the original spirit of speaker feature learning.

Our group follows the feature learning scheme. Recently, we proposed a convolutional time-delay deep neural network
structure (CT-DNN) that can learn speaker features very well~\cite{li2017deep}, and the feature-based SV
outperformed the i-vector model especially in short utterance conditions. This paper extends this work and
investigates the performance of the feature-based SV in situations with language mismatch.


\section{Deep feature learning structure}
\label{sec:model}

This section presents our DNN structures for deep speaker feature learning.
Firstly we review the basic phone-blind CT-DNN structure proposed in~\cite{li2017deep},
and then propose a phone-aware structure that employs phonetic information as
a conditional variable to regularize the speaker feature learning.

\subsection{Phone-blind structure}

\begin{figure*}[htp]
\centering
\includegraphics[width=0.95\linewidth]{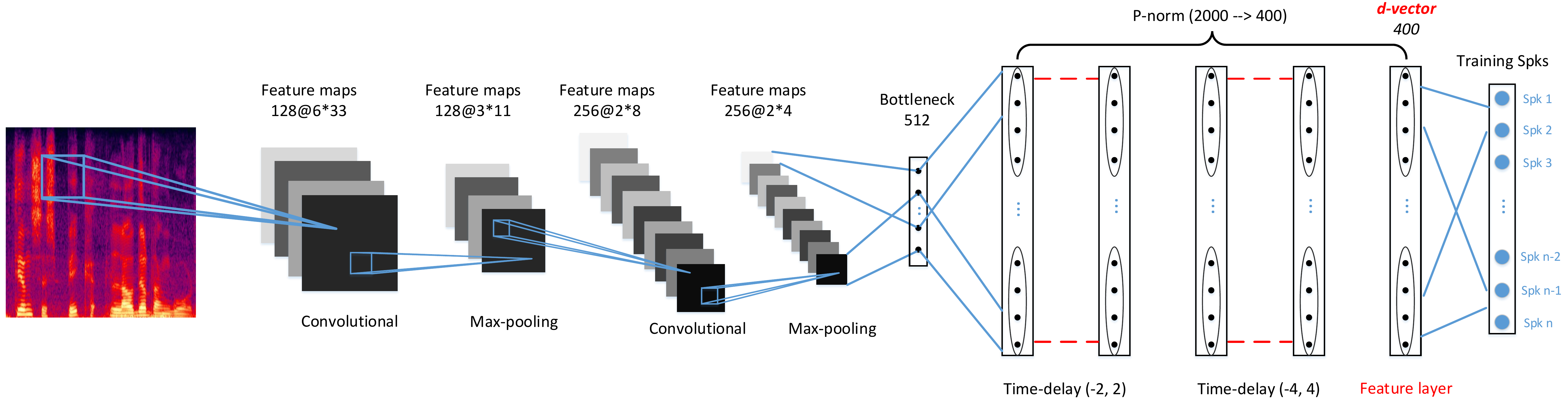}
\caption{The basic deep feature learning structure.}
\label{fig:ctdnn}
\end{figure*}

The basic DNN feature learning structure is illustrated in Fig.~\ref{fig:ctdnn}.
This structure, denoted by `CT-DNN', involves several convolutional layers to extract local
discriminative patterns from the raw features, and several time-delayed layers to increase the effective
temporal context. More specifically, it consists of a convolutional (CN) component and a time-delay (TD) component, connected
by a bottleneck layer consisting of $512$ hidden units.
The convolutional component involves two CN layers, each followed by a max-pooling layer.
This component is used to learn local patterns that are useful in representing speaker traits.
The TD component involves two TD layers, each followed by a P-norm layer.
This component is used to extend the temporal context.
The settings for the two components, including the patch size, the number of feature maps, the
time-delay window, the group size of the P-norm, have been shown in Fig.~\ref{fig:ctdnn}.
A simple calculation shows that with these settings, the size of the effective context window is $20$ frames.
The output of the P-norm layer is projected to a feature layer consisting of $400$ units, where
the vector of the units are re-normalized to a fixed length $400$. This $400$-dimensional normalized
vector is used as the deep speaker feature.
This feature layer is finally connected to the output layer whose units correspond to the speakers in the training data.

\begin{figure*}[htp]
\centering
\includegraphics[width=0.95\linewidth]{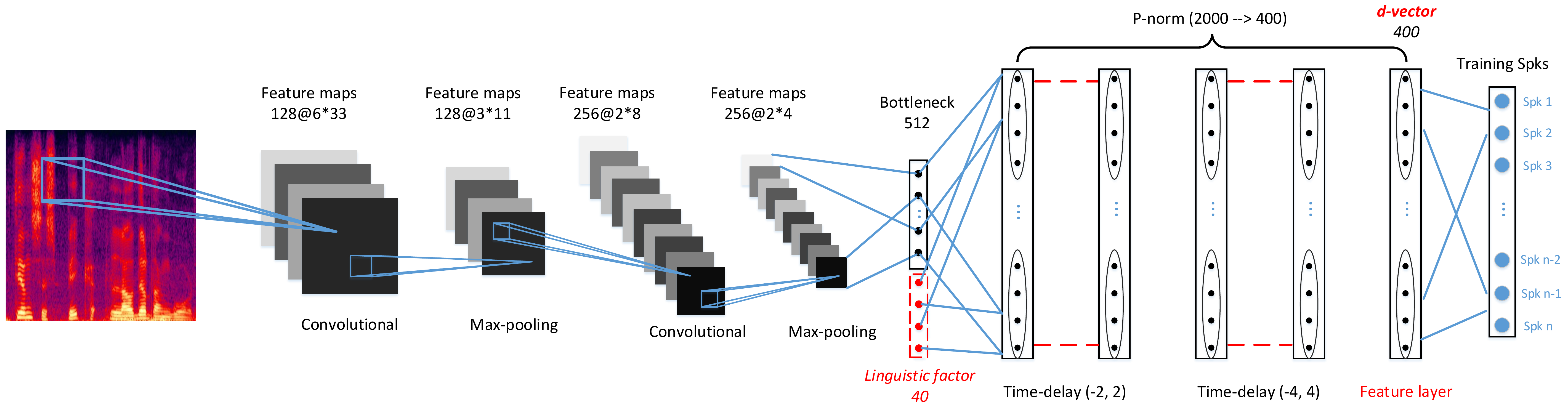}
\caption{The phone-aware deep feature learning structure.}
\label{fig:phone}
\end{figure*}

\subsection{Phone-aware structure}

A potential problem of the CT-DNN model described in the previous section is that it is a `blind learning', i.e., the features
are learned from raw data without prior information. This blind has to deal with the large within-speaker variations caused by
the phonetic content. This is a challenging task and requires more complex models and more speech data.
A possible improvement is to inform the CT-DNN model the phonetic content of each frame, by which most of the within-speaker
variations can be explained away, hence simplifying the speaker feature learning.
This leads to a phone-aware CT-DNN architecture, as shown in Fig.~\ref{fig:phone}, where linguistic factor represents the
phonetic content information. The phonetic content information, or the linguistic factor, can be produced by any model that
can discriminate phones. In this study, we use a DNN model that has been well trained for Chinese speech recognition.

The two deep feature learning structures can be trained using the natural stochastic gradient
descent (NSGD)~\cite{povey2014parallel} algorithm.
Once the models have been trained, the deep speaker feature can be read from the feature layer,
i.e., the last hidden layer of the models.

\section{Experiments}
\label{sec:exp}

In this section, we first present the database used in the experiments, then report the results with
the i-vector and the two feature-based systems, one is phone-blind and the other is phone-aware.
All the experiments were conducted using the Kaldi toolkit~\cite{povey2011kaldi}.

\subsection{Database}
\label{sec:data}

The \emph{Fisher} database and the \emph{CSLT-CUDGT2014 Chinese-Uyghur bilingual} database were used in our experiments.
All the database is in 8kHz sampling rate.
The training set and the evaluation set are presented as follows.

\begin{itemize}
    \item \textbf{Training set}:
    It consists of $2,500$ male and $2,500$ female speakers, with $95,167$ utterances randomly selected from the \emph{Fisher} database,
    and each speaker has about $20$ utterances and totally $120$ seconds in length.
    This dataset was used for training the i-vector system, LDA model, PLDA model, and two deep speaker systems.
    \item \textbf{Evaluation set}: The CSLT-CUDGT2014 Chinese-Uyghur bilingual database. It consists of $181$ speakers, each speaking $10$ Chinese digital strings and
    $10$ Uyghur digital strings. Each string contains $8$ Chinese or Uyghur digits, and is about $2$-$3$ seconds.

\end{itemize}

The test were conducted in $3$ conditions, as shown in Table~\ref{tab:data}.
`LNG1-LNG2' means that enroll with utterances in language LNG1 and test with utterances in language LNG2.
As the trials are symmetric pairs, the `CHS-UYG' and `UYG-CHS' share the same data profile.
We highlight that this data profile and test setting involve highly complex
cross-lingual effect: the training is in English, and the enrollment and test can be in either
Chinese or Uyghur. This is therefore a challenging benchmark to evaluate SV systems
in cross-lingual situations.

\begin{table}[htp]
    \begin{center}
        \caption{Data profile of the test conditions.}
        \label{tab:data}
        \scriptsize
          \begin{tabular}{|l|c|c|c|}
           \hline
               Test condition            &   CHS-CHS    &  UYG-UYG  &  CHS-UYG/UYG-CHS \\
           \hline
           \hline
               \#. of Utts.              &   1,779     &   1,779    &   3,558   \\
           \hline
               Avg. dur of Utts.         &   2.20s     &    2.50s   &  2.35s    \\
           \hline
                \#. of Target trials     &  7.87k      &   7.87k    &   17.52k     \\
                \#. of Nontarget trials  &  1.57M      &   1.57M    &   3.15M       \\
           \hline
          \end{tabular}
    \end{center}
\end{table}

\subsection{Model settings}

We built an i-vector system as the baseline.
The raw feature involved $19$-dimensional MFCCs plus the log energy.
This raw feature was augmented by its first- and second-order derivatives, resulting in a
60-dimensional feature vector. This feature was used by the i-vector model.
The UBM was composed of $2,048$ Gaussian components, and the dimensionality of the i-vector space was $400$.
The dimensionality of the LDA projection space was set to $150$.
Prior to PLDA scoring~\cite{Ioffe06}, i-vectors were centered and length normalized. The entire
system was trained using the Kaldi SRE08 recipe.

For the phone-blind d-vector system, the architecture was based on Fig.~\ref{fig:ctdnn}.
The input feature was 40-dimensional Fbanks, with a symmetric $4$-frame window to splice the neighboring frames, resulting in $9$ frames in total.
The number of output units was $5,000$, corresponding to the number of speakers in the training data.
The speaker features were extracted from the last hidden layer (the feature layer in Figure~\ref{fig:ctdnn}),
and the utterance-level d-vectors were derived by averaging the frame-level features.
The scoring metrics used for the i-vector system were also used for the d-vector system during the test, including cosine distance, LDA and PLDA.

For the phone-aware d-vector system, the phonetic DNN was trained for Chinese speech recognition (ASR).
The training used more than $1,400$ hours Chinese telephone speech, using the Kaldi toolkit following the WSJ nnet3 s5 recipe.
The input feature was 40-dimensional Fbanks. With $7$ time-delay hidden layers, the valid context window of each frame was $37$ frames.
Each hidden layer contained $1,024$ hidden units, and the output layer contained $3,509$ units, corresponding to the number of GMM senones.
To produce the linguistic factor for the phone-aware CT-DNN, a Singular Value Decomposition (SVD) was applied to decompose
the final affine transformation matrix of the ASR DNN, by setting the rank of the SVD to $40$. The $40$-dimensional activations were read from the low-rank layer of the decomposed matrix, and were used as the linguistic factor of the CT-DNN model.

\subsection{Experimental results}

    \begin{table}[htb]
    \begin{center}
      \caption{The EER(\%) results of cross-lingual speaker verification.}
      \label{tab:clng}
          \begin{tabular}{|c|l|c|c|c|}
            \hline
            \multicolumn{2}{|c|}{}               & \multicolumn{3}{|c|}{EER\%}\\
            \hline
               Systems              &  Metric    &   CHS-CHS  &  UYG-UYG &  CHS/UYG \\
           \hline
           \hline
               i-vector             &    Cosine  &   7.55    &   6.16   &   15.14   \\
                                    &    LDA     &   6.30    &   5.63   &   12.77   \\
                                    &    PLDA    &   5.31    &   4.29   &   9.82    \\
            \hline
               d-vector             &    Cosine  &   4.17    &   4.09   &   10.45   \\
               (phone-blind)        &    LDA     &   6.64    &   5.47   &   13.16   \\
                                    &    PLDA    &   3.75    &   3.71   &   8.66    \\
            \hline
               d-vector             &    Cosine  &   4.07   &   4.03    &  10.30  \\
              (phone-aware)         &    LDA     &   6.09   &   5.21    &  13.02  \\
                                    &    PLDA    &\textbf{3.61}  &\textbf{3.52}    &\textbf{8.37} \\
           \hline
          \end{tabular}
      \end{center}
   \end{table}

The results are shown in Table~\ref{tab:clng}. As `CHS-UYG' and `UYG-CHS' are symmetric in the test, they are merge as `CHS/UYG'.
From these results, we can observe that with all the three systems, the models trained with the Fisher database
still work on the new dataset CSLT-CUDGT2014, it is in totally different languages.
This indicates that both the i-vector system and d-vector system posses certain cross-lingual generalizability.
However, the d-vector systems outperform the i-vector system with a large margin. According to our previous work~\cite{li2017deep},
in the language-matched test, the i-vector system has no such big advantage on the $3$-seconds test condition. This
therefore demonstrated that the deep feature
systems are more robust against language mismatch.
Compared to CHS-CHS and UYG-UYG, the performance with enrollment-test language mismatch (CHS/UYG)
is clearly worse, though the performance with the d-vector systems is clearly superior compared to the i-vector system.

Compared with the phone-blind d-vector system, the phone-aware d-vector system exhibits better performance.
This demonstrated that adding the phonetic information can reduce the burden of the CT-DNN model in
learning the phonetic-relevant variation,  hence producing speaker features with better quality.

The robustness of the d-vector systems in situation with language mismatch demonstrated that the deep features
learned by the CT-DNNs have caught some `fundamental patterns' of speakers, and eliminated most language-relevant
variations. This is a strong evidence that the deep features are highly robust and generalizable, which is
the key value that deep learning offers to the SV research.

\section{Conclusions}
\label{sec:conl}

This paper investigated the feature-based speaker verification approach in situations with language mismatch between training,
enrollment and test.
Two deep feature learning structures were studied, one is phone-blind and the other is phone-aware. The
experimental results demonstrated that the feature-based SV systems work very well in cross-lingual situations
and outperformed the state-of-the-art i-vector/PLDA system, and the phone-aware system is more superior.
This indicates that the proposed feature learning model indeed learned how to extract the `fundamental patterns' of speakers,
and the extracted features are robust and generalizable.
In the further work, we will investigate the robustness and generalizability of deep speaker features in other situations, e.g.,
with channel mismatch and strong background noise. More powerful feature learning models will be investigated as well.

\section*{Acknowledgment}

This work was supported by the National Natural Science Foundation of China under Grant No. 61371136 / 61633013
and the National Basic Research Program (973 Program) of China under Grant No. 2013CB329302.

\bibliographystyle{IEEEtran}
\bibliography{mybib}

\begin{thebibliography}{10}
\providecommand{\url}[1]{#1}
\csname url@samestyle\endcsname
\providecommand{\newblock}{\relax}
\providecommand{\bibinfo}[2]{#2}
\providecommand{\BIBentrySTDinterwordspacing}{\spaceskip=0pt\relax}
\providecommand{\BIBentryALTinterwordstretchfactor}{4}
\providecommand{\BIBentryALTinterwordspacing}{\spaceskip=\fontdimen2\font plus
\BIBentryALTinterwordstretchfactor\fontdimen3\font minus
  \fontdimen4\font\relax}
\providecommand{\BIBforeignlanguage}[2]{{%
\expandafter\ifx\csname l@#1\endcsname\relax
\typeout{** WARNING: IEEEtran.bst: No hyphenation pattern has been}%
\typeout{** loaded for the language `#1'. Using the pattern for}%
\typeout{** the default language instead.}%
\else
\language=\csname l@#1\endcsname
\fi
#2}}
\providecommand{\BIBdecl}{\relax}
\BIBdecl

\bibitem{campbell1997speaker}
J.~P. Campbell, ``Speaker recognition: A tutorial,'' \emph{Proceedings of the
  IEEE}, vol.~85, no.~9, pp. 1437--1462, 1997.

\bibitem{reynolds2002overview}
D.~A. Reynolds, ``An overview of automatic speaker recognition technology,'' in
  \emph{Acoustics, speech, and signal processing (ICASSP), 2002 IEEE
  international conference on}, vol.~4.\hskip 1em plus 0.5em minus 0.4em\relax
  IEEE, 2002, pp. IV--4072.

\bibitem{kinnunen2010overview}
T.~Kinnunen and H.~Li, ``An overview of text-independent speaker recognition:
  From features to supervectors,'' \emph{Speech communication}, vol.~52, no.~1,
  pp. 12--40, 2010.

\bibitem{hansen2015speaker}
J.~H. Hansen and T.~Hasan, ``Speaker recognition by machines and humans: A
  tutorial review,'' \emph{IEEE Signal processing magazine}, vol.~32, no.~6,
  pp. 74--99, 2015.

\bibitem{ma2004english}
B.~Ma and H.~Meng, ``English-chinese bilingual text-independent speaker
  verification,'' in \emph{Acoustics, Speech, and Signal Processing, 2004.
  Proceedings.(ICASSP'04). IEEE International Conference on}, vol.~5.\hskip 1em
  plus 0.5em minus 0.4em\relax IEEE, 2004, pp. V--293.

\bibitem{auckenthaler2001language}
R.~Auckenthaler, M.~J. Carey, and J.~S. Mason, ``Language dependency in
  text-independent speaker verification,'' in \emph{Acoustics, Speech, and
  Signal Processing, 2001. Proceedings.(ICASSP'01). 2001 IEEE International
  Conference on}, vol.~1.\hskip 1em plus 0.5em minus 0.4em\relax IEEE, 2001,
  pp. 441--444.

\bibitem{Abhinav}
A.~Misra and J.~H.~L. Hansen, ``Spoken language mismatch in speaker
  verification: An investigation with nist-sre and crss bi-ling corpora,'' in
  \emph{IEEE Spoken Language Technology Workshop}.\hskip 1em plus 0.5em minus
  0.4em\relax IEEE, 2014, pp. 372--377.

\bibitem{askar2016}
A.~Rozi, D.~Wang, L.~Li, and T.~F. Zheng, ``Language-aware {PLDA} for
  multilingual speaker recognition,'' in \emph{O-COCOSDA 2016}, 2016.

\bibitem{Reynolds00}
D.~Reynolds, T.~Quatieri, and R.~Dunn, ``Speaker verification using adapted
  gaussian mixture models,'' \emph{Digital Signal Processing}, vol.~10, no.~1,
  pp. 19--41, 2000.

\bibitem{Kenny07}
P.~Kenny, G.~Boulianne, P.~Ouellet, and P.~Dumouchel, ``Joint factor analysis
  versus eigenchannels in speaker recognition,'' \emph{IEEE Transactions on
  Audio, Speech, and Language Processing}, vol.~15, pp. 1435--1447, 2007.

\bibitem{dehak2011front}
N.~Dehak, P.~J. Kenny, R.~Dehak, P.~Dumouchel, and P.~Ouellet, ``Front-end
  factor analysis for speaker verification,'' \emph{IEEE Transactions on Audio,
  Speech, and Language Processing}, vol.~19, no.~4, pp. 788--798, 2011.

\bibitem{Kinnunen10}
T.~Kinnunen and H.~Li, ``An overview of text-independent speaker recognition:
  From features to supervectors,'' \emph{Speech communication}, vol.~52, no.~1,
  pp. 12--40, 2010.

\bibitem{wang2013vocal}
J.~Wang and M.~T. Johnson, ``Vocal source features for bilingual speaker
  identification,'' in \emph{Signal and Information Processing (ChinaSIP), 2013
  IEEE China Summit \& International Conference on}.\hskip 1em plus 0.5em minus
  0.4em\relax IEEE, 2013, pp. 170--173.

\bibitem{li2017deep}
L.~Li, Y.~Chen, Y.~Shi, Z.~Tang, and D.~Wang, ``Deep speaker feature learning
  for text-independent speaker verification,'' \emph{arXiv preprint
  arXiv:1705.03670}, 2017.

\bibitem{akbacak2007language}
M.~Akbacak and J.~H. Hansen, ``Language normalization for bilingual speaker
  recognition systems,'' in \emph{Acoustics, Speech and Signal Processing,
  2007. ICASSP 2007. IEEE International Conference on}, vol.~4.\hskip 1em plus
  0.5em minus 0.4em\relax IEEE, 2007, pp. IV--257.

\bibitem{askar2015cross}
R.~Askar, D.~Wang, F.~Bie, J.~Wang, and T.~F. Zheng, ``Cross-lingual speaker
  verification based on linear transform,'' in \emph{Signal and Information
  Processing (ChinaSIP), 2015 IEEE China Summit and International Conference
  on}.\hskip 1em plus 0.5em minus 0.4em\relax IEEE, 2015, pp. 519--523.

\bibitem{lu2009effect}
L.~Lu, Y.~Dong, X.~Zhao, J.~Liu, and H.~Wang, ``The effect of language factors
  for robust speaker recognition,'' in \emph{Acoustics, Speech and Signal
  Processing, 2009. ICASSP 2009. IEEE International Conference on}.\hskip 1em
  plus 0.5em minus 0.4em\relax IEEE, 2009, pp. 4217--4220.

\bibitem{ehsan14}
V.~Ehsan, L.~Xin, M.~Erik, L.~M. Ignacio, and G.-D. Javier, ``Deep neural
  networks for small footprint text-dependent speaker verification,'' in
  \emph{Acoustics, Speech and Signal Processing (ICASSP), 2014 IEEE
  International Conference on}, vol.~28, no.~4, 2014, pp. 357--366.

\bibitem{heigold2016end}
G.~Heigold, I.~Moreno, S.~Bengio, and N.~Shazeer, ``End-to-end text-dependent
  speaker verification,'' in \emph{Acoustics, Speech and Signal Processing
  (ICASSP), 2016 IEEE International Conference on}.\hskip 1em plus 0.5em minus
  0.4em\relax IEEE, 2016, pp. 5115--5119.

\bibitem{zhang2017end}
S.-X. Zhang, Z.~Chen, Y.~Zhao, J.~Li, and Y.~Gong, ``End-to-end attention based
  text-dependent speaker verification,'' \emph{arXiv preprint
  arXiv:1701.00562}, 2017.

\bibitem{snyderdeep16}
D.~Snyder, P.~Ghahremani, D.~Povey, D.~Garcia-Romero, Y.~Carmiel, and
  S.~Khudanpur, ``Deep neural network-based speaker embeddings for end-to-end
  speaker verification,'' in \emph{SLT'2016}, 2016.

\bibitem{li2017}
C.~Li, X.~Ma, B.~Jiang, X.~Li, X.~Zhang, X.~Liu, Y.~Cao, A.~Kannan, and Z.~Zhu,
  ``Deep speaker: an end-to-end neural speaker embedding system,'' \emph{arXiv
  preprint arXiv:1705.02304}, 2017.

\bibitem{povey2014parallel}
D.~Povey, X.~Zhang, and S.~Khudanpur, ``Parallel training of dnns with natural
  gradient and parameter averaging,'' \emph{arXiv preprint arXiv:1410.7455},
  2014.

\bibitem{povey2011kaldi}
D.~Povey, A.~Ghoshal, G.~Boulianne, L.~Burget, O.~Glembek, N.~Goel,
  M.~Hannemann, P.~Motlicek, Y.~Qian, P.~Schwarz \emph{et~al.}, ``The kaldi
  speech recognition toolkit,'' in \emph{IEEE 2011 workshop on automatic speech
  recognition and understanding}, no. EPFL-CONF-192584.\hskip 1em plus 0.5em
  minus 0.4em\relax IEEE Signal Processing Society, 2011.

\bibitem{Ioffe06}
S.~Ioffe, ``Probabilistic linear discriminant analysis,'' \emph{Computer Vision
  ECCV 2006, Springer Berlin Heidelberg}, pp. 531--542, 2006.

\end{thebibliography}

\end{document}